\documentclass[12pt]{article}

\usepackage{epsfig}

\usepackage{scicite}

\usepackage{times}

\newenvironment{sciabstract}{%
\begin{quote} \bf }
{\end{quote}}

\newcounter{lastnote}
\newenvironment{scilastnote}{%
\setcounter{lastnote}{\value{enumiv}}%
\addtocounter{lastnote}{+1}%
\begin{list}%
{\arabic{lastnote}.}
{\setlength{\leftmargin}{.22in}}
{\setlength{\labelsep}{.5em}}}
{\end{list}}

\title{Currents induced by  domain wall motion in thin ferromagnetic wires}

\author
{S. E. Barnes,$^{1,2\ast}$ S. Maekawa$^{1}$\\
\\
\normalsize{$^{1}$Institute for Materials Research, Tohoku University Sendai 980-8577, Japan}\\
\normalsize{$^{2}$Physics Department, University of Miami, Coral Gables, FL 33124, USA
}\\
\\
\normalsize{$^\ast$To whom correspondence should be addressed; E-mail:  barnes@physics.miami.edu.}
}

\date{}

\begin{document} 


\baselineskip24pt


\maketitle


\begin{sciabstract}
Historically the role of eddy currents in the motion of  Bloch walls has been extensively studied. In contrast the direct interaction of currents with such a moving wall has received very little attention. However recently it has been shown that a wall can be displaced by the effects of a current alone, a fact which is important to the development of spintronics. We report here that such a moving wall within an itinerant ferromagnet constitutes a seat of electro-motive-force (or more accurately a spin-motive-force) which  induces a current  flow in an external circuit. While an external magnetic field is the most evident means by which to induce Bloch wall motion, it is not the only possibility and the forces intrinsic to certain geometries imply that the state of a spintronics memory device might be read via this mechanism using currents alone and that power amplification of spin polarized pulses is possible. An estimate of the induced current shows it to be technologically useful.
\end{sciabstract}



\textwidth 5.4truecm 
Spintronic devices have great technological promise but represent a challenging problem at both an applied and fundamental level. It has been  shown both theoretically\cite{1,2} and experimentally\cite{3,4} that a Bloch wall might be driven down a nano-sized wire through the effects of currents alone. Using the fact that such a wall can be pinned at some chosen point in the wire, this effect can be used, e.g.,  to write bits into a memory element. It is usually envisaged that the state of such a device be read using the GMR effect\cite{4}. We  demonstrate here,  when a Bloch wall moves, it induces  a spin current and hence, for spin polarized conduction electrons, constitutes a seat of electro-motive-force (emf). More accurately the moving wall constitutes a spin-motive-force (smf). This is the elusive\cite{5} equivalent of an emf but which drives spin rather than charge currents.  The application of a {\it static\/} magnetic field is the most evident fashion by which to induce Bloch wall motion and it is an interesting curiosity that an emf/smf can be produced by a {\it time independent\/} magnetic field. It is also known\cite{6}  that a closed Bloch wall will tend to collapse under its own surface tension. The fact that a wall will move to reduce its total energy, also implies that in restricted geometries it will suffer a force. We will describe how such effects permit  the state of a memory device to be not only  written but also read {\it directly\/} using uniquely currents. This is of appreciable technological importance.

Corresponding to  Faraday's law, the elementary definition of an emf is  ${\cal E} =  \oint(\vec f/(-e))\cdot d \vec s\,$ where $\vec f = (-e)(\vec E + \vec v \times \vec B)$ is the force on the charge $-e$ carriers. Here $\vec E$ and $\vec B$ are the electric and magnetic fields respectively.  Analyzed here is an unappreciated additional source of an emf. Again it is an elementary fact that, e.g., in a Stern-Gerlach experiment, there is a force $\vec f_s = - ({\partial /\partial \vec r})\,\vec \mu \cdot \vec B,$
where $\vec \mu$ is the spin magnetic moment. In the current situation $\vec f_s$ is due to a finite ${\partial \vec \mu /\partial \vec r}$ rather than a ${\partial \vec B /\partial \vec r}$ and, in particular, occurs  because $\langle \vec \mu \rangle$ changes direction within a Bloch wall. Evidently  $\vec f_s$ should be included in  the definition of ${\cal E} $.

\begin{figure}[t!]\centerline{\epsfig{file=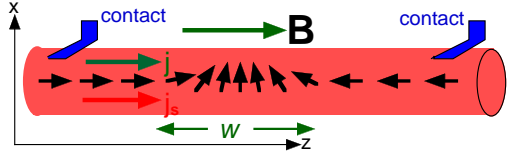,width=2.8in}  }
\caption[toto]{Far from the central wall the spins point along the easy axis taken to be the $z$-direction.  There is also a magnetic field $\vec B$ in the $z$-direction. The wall lies in the $x-z$-plane. The conduction electrons carry  spin $j_s$ {\it and\/} charge currents $j$. }
\label{F0}\end{figure}

The central results are first deduced using conservation principles. The  wire geometry is shown in Fig.~\ref{F0}. The anisotropy energy is such that the magnetization far from the Bloch wall points along the wire and this is taken to be the $z$-direction. In order to exert a force on the wall, a static  magnetic field $\vec B$ is applied along the same direction. For an itinerant ferromagnet, in addition to the local-magnetic-moments which constitute the wall, are conduction electrons which can carry a current. It is assumed initially  that there is conservation of, $J_z$, the  $z$-component  of the {\it total\/} angular momentum.  The motion of the wall constitutes a local-moment spin current and the conservation of $J_z$ requires an equal and opposite conduction electron spin current. If $A_0$ is the cross-sectional area, $v_c$ is unit cell volume, $S$ the net spin per site, and $v$ its velocity, then in a time $\Delta t$ there are $v \Delta t A_0/v_c$ spins which effectively move towards the wall to constitute a spin current $j_S = Sv/v_c$. Unlike charge currents, in general, conduction electron spin currents are {\it not\/} conserved. However, in an itinerant ferromagnet, the presence of a charge current $j_e$ implies a spin current $j_s = p(j_e/2e )$, where $p$ is the effective magnetic moment per charge carrier.  Like the magnetic polarization $m$, $p$ is a material specific parameter with $p\sim 0.7$ for Permalloy\cite{3}. As such, it is a property of the current carrying (thermodynamic) ground state and the relationship  $j_s = p(j_e/2e )$ will apply provided that the conduction electrons pass through the wall adiabatically, which must be the case for a large enough width $w$, Fig.~1. Assuming this to be the case and equating the two spin currents,
$$
j_e =  {2e S\over  p v_c}v,
\eqno(1)
$$
gives a relationship between the charge current and wall velocity which is well known\cite{3}. Our assumption of the conservation of angular momentum implies that the wall does not move in the absence of a current, i.e., that there is a coupling between the two sub-systems which will be expanded upon below. This interaction must also conserve energy. The energy decrease of the wall as it moves must be compensated by an increase in that of the conduction electrons, and which is dissipated in the load. Given that $S g\mu_B B$, where $g$ is the $g$-factor and  $\mu_B$ is the Bohr magnetron, is the Zeeman energy, the local moment energy  change, in the time $\Delta t$ is $2S g\mu_B B v \Delta t A/v_c$ while the energy dissipated by  the current is ${\cal E} j_e A \Delta t$, to give for the emf,
$$
{\cal E} = {p g \mu_B B \over e}.
\eqno(2)
$$
If $R$ is the load resistance then $j_e= I/A_0 = {\cal E}/A_0 R$ is determined by  the emf, i.e., the wall velocity reflects the load resistance. (Ignoring  eddy current and relaxation losses.)

The  Eqs.~1 and 2 along with their applications to basic devices, detailed below, are the principal results of this Report. However, it is the case these equations result from rather sweeping assumptions about the conservation of angular momentum and energy and which are easily questioned. We have developed a more substantial theory at  two levels. What is described in the following amounts to a macroscopic theory. Based upon half-metal limit of the strongly correlated double exchange model, a fully microscopic approach has also been developed. These calculations are reported in the supplementary material in a methods section and substantiate the macroscopic considerations. There is no want of existing theories which might be applicable to this spin transfer problem but none is, in fact, adequate since they cannot address, in a non-perturbative fashion, the two issues behind Eqs.~1-2, i.e., the simultaneous conservation of energy and angular momentum. 

It is a fact, which will be exposed below, that the basic coupling, which conserves the angular momentum and energy, is non-perturbative, i.e., does not arise from the interaction which creates the internal exchange field but rather has as its origin the conduction electron kinetic energy. It arises when certain quantum corrections are added to the classical description of the spin degrees of freedom. Magnons are the quantum excitations of a regular ferromagnet. The wavelength $\vec q =0$ such magnon is a Goldstone boson which restores the broken symmetry of a ferromagnet. When a Boch wall is present, for an infinite uniform system without fields, there is an additional broken symmetry corresponding to specifying the position $z_0$ of the center of the wall.  The restoration of this symmetry is essential to the aforementioned conservation principles. There is a new Goldstone boson $b^\dagger_w$, defined below, which has this function. For this reason it is necessary to make a digression into the theory of magnons. As is usual\cite{7}, the Holstein-Primakoff transformation $S_{iz}=\hbar(S-b^\dagger_i b^{}_i)$, $S^+_i = \hbar (2S- b^\dagger_i b^{}_i)^{1/2}b^{}_i \approx \hbar (2S)^{1/2}b^{}_i$ is used to quantize the local moment spin, $\vec S_i$,  for site $i$. In principle, the axes of quantization are arbitrary, {\it however\/}, as the standard theory\cite{7} of ferromagnetic magnons well illustrates, {\it in fact\/}, if the magnons are to all have positive energies, the local axis of quantization must be taken along the {\it classical\/} equilibrium direction, i.e., the direction appropriate in the limit $S\to \infty$. This  local equilibrium direction, for site $i$, is specified by the Euler angles $\theta_i$ and $\phi_i$.

The effective Hamiltonian is then, ${\cal H} \approx {\cal V}\{\theta_i, \phi_i\} + {\cal H}_t + {\cal H}_s + {\cal H}_{ts}$, where the ($B=0$) classical energy functional ${\cal V}\{\theta_i, \phi_i\} = -S^2 J \sum_{<ij>}\cos (\theta_i-\theta_j) -S^2A\sum_i \cos^2 \theta_i $, involves $J$  and $A$ which are  the exchange and anisotropy {\it energy\/} constants, where the sum $\sum_{<ij>}$ is over near-neighbors.  The kinetic energy of the conduction electrons is  ${\cal H}_t$, while ${\cal H}_s$ and ${\cal H}_{ts}$ are  the magnon and interaction Hamiltonians\cite{2}. The Bloch wall structure is determined by minimizing this ${\cal V}\{\theta_i, \phi_i\}$ and yields the usual result, $\theta(z) =2\cot^{-1} e^{-((z-z_0)/w)}$, with $\phi_i =0$, for a wall centered at $z_0$, where the wall width $ w= a\left({J \over 2A}\right)^{1/2}$. Here $a$ is the lattice spacing in the $z$-direction. The local axes used to quantize the local moments changes from $\theta = 0$ to $\theta = \pi$ with increasing $z$ and the wall lies in the $x-z$-plane, Fig.~1. 

A second digression is needed in order to explain the role of the adiabatic theorem. In a discrete  model, if $t$ is the hopping matrix element, and if the internal exchange field $J \gg t$, corresponding to a half metal, then it is possible to treat $t$ as a perturbation. Clearly the maximum of exchange energy of a plane corresponds to a maximum in its angular momentum and excited states have energies $\sim J$. In the limit $J \gg t$,  electron hopping matrix elements  adiabatically connect these exchange ground states, and for each such ground state the conduction electron spin $\vec s$ is aligned parallel to the total spin $\vec S$ (per site). However when the angles between the angular momentum vectors of adjacent planes are small, as in a Bloch wall, the electrons see an almost constant exchange field and the above inequality is certainly too strong. The conduction electron spins rotate by $\pi$ as they pass through the wall of width $w$.  They must therefore rotate by an angle $\sim (a/w)\pi$ per hopping event. These take a time $\sim \hbar/t$ and therefore  the rate of rotation is $\omega_r \sim (t/\hbar)(a/w)\pi$. In the frame rotating with the electrons, there is an effective transverse field $\hbar \omega_r  /g \mu_B$\cite{8}. These same electrons experience a longitudinal internal exchange field $SJ/g \mu_B$ and spin-flip transitions are negligible, i.e., the adiabatic theorem is satisfied, if  $SJ > \pi (a/w) t$ a condition which is well satisfied  for, e.g., Permalloy, with $S\sim 0.5$, $J\sim 0.3$eV, $t\sim 1-2$eV, $w\sim 200$nm and $a\sim 3$\AA. Since the majority and minority spins are oppositely aligned, the adiabatic condition dictates that
 $$
 \vec s  = \pm  {\vec S \over 2S} 
 \eqno(3)
$$
with the upper (lower) sign  for the majority (minority) electron spin direction.  Since majority and minority electrons rotate into themselves, their populations are invariant and Eq.~3 correctly implies $\langle \vec s \rangle = m \vec S/2S$\cite{9}. The adiabatic transport of the (thermodynamic)  ground state through the wall also implies a {\it fixed\/} relationship between the spin, $j_s$,  and charge currents, $j_e$. Far from the wall this ground state is such that $j_s = p j_e/2e$, where $p$ is the material dependent constant mentioned above. It is this state which is transported adiabatically through the wall and as a consequence this relationship remains valid even within the wall.

The sought for coupling arises when Schr\"odinger's equation is written in the frame which rotates with the wall. In this frame there must  be a ${\cal H}_{ts} = (1/2)\int dV (\vec A \cdot \vec j_e)$ coupling of an appropriate ($e=c=1$) vector potential, $\vec A$, to the particle  current $\vec j_e$. This follows since the transformation to the local axes of quantization within the wall corresponds to a gauge transformation, i.e., if $\psi(z, t)$ is the conduction electron wave function, in the continuum limit, the rotation used to generate a stationary wall, causes $\psi  (z, t) \to e^{i\lambda(z)}\psi  (z, t)$, where $\lambda (z) = \theta(z) s_{y}(z) /\hbar $. With $f(z,t) = \hbar \lambda (z) $, the usual transformations of the potentials are  $\vec A \to \vec A  - \vec \nabla f(z)$ and $\phi \to \phi + (\partial f /\partial t) $,  i.e., $\phi \to \phi$ and, in discrete notion,  the finite component $A_{iz} =  (\partial \theta_i /\partial z) s_{iy}$. Using Eq.~3 to make the replacement $s_{iy} \to \pm S_{iy}/2S$ with the upper (lower) sign for the majority (minority) spins, the discrete equivalent of $(1/2)\int dV (\vec A \cdot \vec j_n)$ becomes,
$$
{\cal H}_{ts} = 
p {   j_e v_c \over 2  e S}
  \sum_i { \partial \theta_i \over \partial z}   \,S_{iy},
\eqno(4)
$$
which, because of the spin dependent sign in Eq.~3, is proportional to the {\it spin\/} current $j_s = p j_e/2e$ where $j_e$ is the usual {\it charge\/} current. This key  result for $p=1$, has been  derived from the double exchange model, as described in the methods section. This is also consistent with the Slonczewski\cite{1} coupling proportional to $j\vec S_i\times (\vec S_{i+1} \times \vec S_i )$ where $i$ and $i+1$ are adjacent sites in the direction of the current flow and with the approach of Bazaliy et al\cite{B}.

In order to correctly describe the motion of a Bloch wall it is necessary to first identify its center of mass momentum $p_z$ and hence its position operator $\hat z_0$. These involve the Goldstone boson $b^\dagger_w$, mentioned above, and which is defined by considering  a simple infinite wire with translational invariance, i.e.,  by definition, without any applied fields or forces. The momentum is the generator of displacements. The product of small $y$-axis rotations $R(\Delta z)=\prod_i \exp ( i (\theta_i - \theta_{i+1})(\Delta z/a)(S_{iy}/\hbar))\approx  (1+ (1/2)\sum_i (\partial \theta_i /\partial z)\Delta z(2S)^{1/2}(b^\dagger_i - b^{}_i))$, produces a translation by $\Delta z$ and the momentum, per atom in a plane of the wall, is identified as  $p_z = i (\hbar /2) (v_c/aA_0) \sum_i (\partial \theta_i /\partial z)(2S)^{1/2} (b^\dagger_i - b^{}_i) $. This defines the Goldstone boson\cite{2}
$$
b^\dagger_w = \left({J \over 8A}\right)^{1/4}{v_c\over aA_0} \sum_i ({\partial \theta_i \over \partial z})a\, b^\dagger_i,
\eqno(5)
$$ 
where the constant of proportionality  reflects the requirement that $[b^{}_w, b^\dagger_w]=1$ and uses the explicit wall solution for $\theta(z)$ given above. Thus $\hat p_z = i \hbar\left({2A \over J}\right)^{1/4} (2S)^{1/2}(1/a)(b^\dagger_w - b^{}_w)$ while the conjugate coordinate, such that $[\hat z_0, \hat p_z]=i\hbar$, is $\hat z_0 = a \left({J \over 2A}\right)^{1/4} (S/2)^{-1/2} (b^\dagger_w + b^{}_w)$.

By definition,  when a pressure $P_z$ acts on the wall, ${\cal H}$ contains  a term,
$$
{\cal H}_p = { P_z v_c \over a} \hat z_0,
\eqno(6)
$$
i.e., a term proportional to the wall coordinate operator $\hat z_0$.    From classical considerations it would be expected that 
$$
P_z = - { a \over  v_c} {\partial {\cal V}\over  \partial z_0} 
\eqno(7)
$$
where  ${\cal V} (z_0)$ is the classical energy {\it per site\/} of a wall at position $z_0$. That Eq.~7 gives the value of $P_z$ appropriate to Eq.~6 follows from the result ${\partial {\cal V} \over \partial z_0} = {i \over \hbar} [{\cal H}, \hat p_z ]$ of footnote\cite{10}. The only relevant term $[{\cal H}_p, \hat p_z ] = { P_z v_c \over a} [\hat z_0, \hat p_z ] = i\hbar { P_z v_c \over a}$ to give Eq.~7. It can be verified directly that a $z$-directed applied field $B$ generates a contribution of the form of Eq.~6  with $P_z = Sg \mu_B B/v_c$ and that this is the classically defined pressure. It is an important observation that a pressure {\it of any origin\/} is equivalent to that due to an applied magnetic field $B$ and so in general the Eq.~2 becomes
$$
{\cal E} = p{P_z v_c \over 2eS},
\eqno(8)
$$
i.e., is determined by the  {\it total\/} pressure $P_z$.

The wall is rendered stationary using two time dependent rotations generated by $r_{i\phi} = e^{iS_{iz} \omega_i t}$ and $r_{i\theta} = e^{iS_{iy} \theta_i (t)}$. The effect of passing to a rotating frame via  $r_{i\phi}$ is to add an effective magnetic field $B_{\omega_i} = \hbar \omega_i/g\mu_B$ directed along the axis of rotation\cite{8}. This is used to eliminate the explicit $z$-axis pressure $P_z$, i.e., 
$$
\hbar \omega_i = \hbar {\partial \phi \over \partial t} = {P_z v_c\over 2S}.
\eqno(9)
$$
This predicts that the wall makes propellor rotations about the $z$ axis. (To this must be added a $z$-axis oscillation when a perpendicular anisotropy is added. The effects of Gilbert relaxation are discussed below.) That there is no static classical solution when $P_z$ is finite follows using the adaptation ${\partial {\cal V} \over \partial \theta_i} = {i \over \hbar} [{\cal H}, (S_{iy}/\hbar) ]$ of footnote\cite{10}. 
Substituting  the operators $\hat z_0$ and $b^\dagger_w$ into Eq.~6 shows ${\cal H}_p = \sum_i c_i (b^\dagger_i+b^{}_i)$, i.e., is {\it linear\/} in the $b^{}_i$. Since $(S_{iy}/\hbar)  = (1/2i)(2S)^{1/2}(b^\dagger_i - b^{}_i)$ it follows that ${\partial {\cal V} \over \partial \theta_i} = (1/\hbar)(2S)^{1/2} c_i \propto P_z$.  A static solution is not possible since this  requires $(\partial {\cal V} /\partial \theta_i) =0$.

The interaction ${\cal H}_{ts}$, which amounts to a field along the $y$-direction, is similarly rendered null via $r_{i\theta} = e^{iS_{iy} \theta_i (t)}$. That the wall is uniformly displaced implies  $\theta_i \equiv \theta(z_i-z_0(t))$ where $z_0(t)$ is the time dependent position of the wall center. Differentiation with respect to time, defining the wall velocity $v\equiv d z_0(t)/dt$, then implies
$$
{\partial \theta_i \over \partial t} = v {\partial \theta_i \over \partial z}.
\eqno(10)
$$
Thus the effective field which results from the time dependent  $r_{i\theta} = e^{iS_{iy} \theta_i (t)}$ produces a term $\propto \hbar v {\partial \theta_i \over \partial z}$ which cancels ${\cal H}_{ts}$ when $v =p ja^3/2eS$. This  agrees with Eq.~2, obtained using the angular momentum conservation  and confirms that ${\cal H}_{ts}$ correctly embodies this conservation principle. 

Traditionally\cite{6} relaxation is reflectd by a Gilbert term $ - (\alpha/S) \vec S_i \times (\partial {\vec S_i}/\partial t)$ which modifies $\vec B_i \to \vec B_i - ( \alpha /Sg\mu_B ) (\partial {\vec S_i}/\partial t)$ the internal  magnetic field seen by spin $i$. However the additional fields {\it move with the wall\/} and cannot contribute to $P_z$, Eq.~7. This leaves Eq.~9 unchanged. In addition  $(\partial {\vec S_i}/\partial t)$ and the new $\vec B_i$ lie  in the  wall plane and Eq.~10 is {\it also\/} unchanged. Dissipation, in addition to that in the load, occurs through forces acting on the electrons and  is reflected by a  reduction of the emf $\cal E$. The Gilbert parameter for Permalloy $\alpha \sim 0.01$ is  the fraction of the energy lost in reversing a single spin. Since $\cal E$ is a similarly defined  work, it is implied that $\cal E$ is reduced very little due to internal relaxation.

It is necessary to confirm Eq.~8 for the emf $\cal E$. The adiabatic condition implies the spin and charge currents have the {\it fixed\/} relationship $j_s = p j_e/2e$ and this  permits Eq.~4 to be viewed as a spin-{\it charge\/} coupling via a vector potential $A^e_z  = - (p/e) (1/2S)(\partial \theta / \partial z) S_{y} $ and which gives the force on a charge carrier via  $f^e_{zs} = - (\partial A^e_z /\partial t) $.  With the wall at the origin, for short times, $S_{y} = \hbar S \sin \theta \sin \omega t + \ldots \approx S \sin \theta  \hbar\omega t$. Thus $f^e_{zs} = (p/e)  (1/2S)(\partial \theta / \partial z)  \sin \theta\, S \hbar \omega = (p/e)  (1/2)(\partial \cos \theta / \partial z)  (P_z v_c/ 2S)$ using Eq.~9.  The  work done is obtained by integrating this.  The result {\it is\/}  Eq.~8.

The predicted current densities $j$ are very large. With $S=1$ and $a\sim 3$\AA\ and a realistic\cite{3} $v \sim 1$m/s the current density $j \sim 3 \times 10^{10}$A/m$^2$ or a current $i \sim 30$mA for a micron square wire. The value of $\cal E$ reflects the value of $B$ and for $1$T corresponding to about 100$\mu$V.

\begin{figure}[t!]\centerline{\epsfig{file=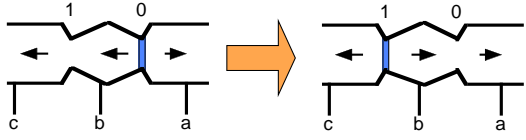,width=2.8in}  }
\caption[toto]{A current only read-write memory element. To the left the system is in the 0-state. A current between $a$ and $b$ will carry the wall past the unstable equilibrium point. As it moves towards 1-state under the force $P_z$ implicit in the device shape it will produce an output emf between $b$ and $c$. No emf occurs if the system is in the 1-state. The system can be switched between 0 and 1 by a passing a current from $a$ to $c$.
}
\label{F1}\end{figure}

A basic memory device is illustrated in Fig.~\ref{F1}. The Boch wall has two stable equilibrium positions $0$ and $1$ and might be switched from one to the other using an external current. In a current only design, the device might be read by applying a current between $a$ and $b$. If the wall lies in position $0$ it will be dislodged by this reading pulse, and once it passes the half-way point, will induce an output in the circuit connecting $b$ and $c$. Clearly, there will be no such current pulse if the wall is initially in position $1$. Here the pressure $P_z$ is produced by the shape of the bridge. (Instead the the device might be read by simply applying a magnetic de-pinning field. The device will only produce a output pulse when the field has the correct sense to de-pin $0$ rather than $1$.) There are any number of other variants too numerous to be described here.

\begin{figure}[t!]\centerline{\epsfig{file=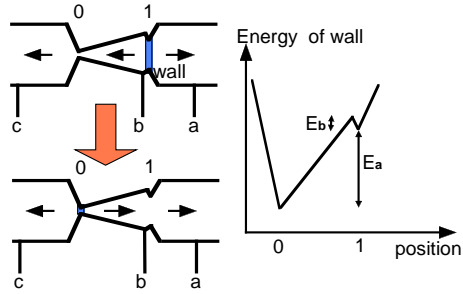,width=2.6in}  } 
\caption[toto]{An power amplifier. Starting with the initial state, top left, a current pulse between $a$ and $b$ moves the wall from 1 to $b$, i.e., a point at which there is a $P_z$ to the left. There is an emf between $b$ and $c$ as the wall moves between $b$ and 0. The final state is shown at the bottom left.  To the right is shown the energy profile of the device, see text.}
\label{F2}\end{figure}

\begin{figure}[t!]\centerline{\epsfig{file=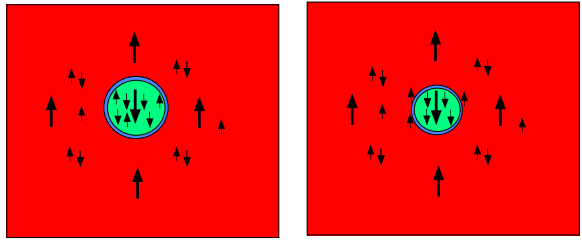,width=3.0in}  } 
\caption[toto]{A spin sieve. The ferromagnetic state is indicated by the large spins. The small spins correspond to the itinerant electrons. The wall shown in blue has an energy proportional to its volume, i.e., there is a surface tension which would cause the bubble to collapse. This finite force $P_z$ causes an smf between the interior and exterior so that the up, but not the down spins flow to the outside. This causes the bubble to get smaller and to have an interior with only down spins. }
\label{F3}\end{figure}

Current and power amplification might be achieved by the device shown in Fig.~\ref{F2}. Now only one of the equilibrium situations $0$ is a truly stable. An initialization pulse  latches the system in the unstable equilibrium position $1$. A small short current pulse between $a$ and $b$ causes the wall to leave $1$ and the large $P_z$ implicit in the design produces an equally large induced emf and hence a large current output in that part of the circuit which is connect to $b$ and $c$. Isolation of input and output is afforded by the small distance 1 to $b$. The potential seen by the wall is also illustrated in Fig.~\ref{F2}.  A large gain implies a small barrier, of height $E_b$, at an energy $E_a$ relative to $0$, and which is large compared to $E_b$. In fact, the power gain $g$ is limited by, and is approximately equal to, 
$
g = {E_a\over E_b},
$ 
since the input pulse must raise the wall over the barrier, i.e., give it an energy $E_a$ while the maximum energy which is given to the external output circuit is evidently $E_a$. In position $1$ the wall might either tunnel or be thermally excited out of this unstable equilibrium, {\it however\/} the wall constitutes a macroscopic object and this affords and enormous reduction in both of these processes. For the sake of illustration imagine the de-pinning field is 0.01T or $\sim 0.01$K. If the wire has $10^3$ atoms in both perpendicular directions there are $10^6$ spins in the cross-section and, for a wall with a length of $w \sim 10^3$ spins, it is implied that the  barrier height is $10^9 \times 0.01\sim 10^7$K which precludes the possibility of direct escape by both thermal excitation and tunneling. 

Finally, there is an interesting field theory aspect to the problem. The aligned local moment ferromagnetic state defines an essentially featureless vacuum. The present system  has two such vacuo connected by a solitonic object, namely the Bloch wall. When acted upon by a force, the wall can {\it only\/} move when there is an appropriate itinerant electron spin current.  In fact, the wall acts as  a ``spin sieve" causing up and down spins to move in opposite directions. In the geometry shown in Fig.~\ref{F3} the wall wishes to collapse under the force of its own surface tension but cannot unless it pumps up spins outwards. A similar situation is possible in which  the magnetic broken symmetry is replaced by that associated with {\it any\/} quantum number. If this was, e.g., the quantum number which distinguishes particles from anti-particles the sieve would separate matter from anti-matter.



\begin{scilastnote}
\item 
We would like to thank Prof. Gerrit Bauer for a critical reading of the manuscript and for a number of constructive comments.
This work was supported by a Grant-in-Ad for Scientific Research on
Priority Areas from the Ministry of Education, Science, Culture and
Technology of Japan, CREST and NAREGI. SEB  wishes to
thank IFCAM and the members of  IMR for their kind hospitality. 
\end{scilastnote}




\clearpage

\noindent{\bf Supplementary material}

\vskip 20pt

\noindent{Methods}

\vskip 15pt

The microscopic theory is based upon a standard model for ferromagnets, i.e.,  the double exchange model: 
$$
{\cal H}
= - t\sum_{<ij>\sigma}\left(c^\dagger_{i\sigma}c^{}_{j\sigma} + H.c.\right) + U\sum_{i}n_{i\uparrow}n_{i\downarrow} 
$$
$$
 - \sum_{i} (J_0  \vec S_i\cdot \vec s_i +A {S_{iz}}^2 
 )
+  \sum_{<ij>}J^0_{ij}
\vec S_i\cdot \vec S_j 
$$
$$
+ 2\mu_B H\sum_{i} (S_{iz} + {1\over 2}\sum_\sigma \sigma c^\dagger_{i\sigma}c^{}_{i\sigma}) -  \mu \hat  N
\eqno(S1)
$$
where $\vec S_i$ is the spin operator and 
$c^\dagger_{i\sigma}$ creates  an electron  with spin $\sigma$ at site $i$.  The wire extends along the $z$-direction which, with $A>0$, is the easy axis. The chemical potential is $ \mu$ and $\hat N$ is the total number operator. For simplicity only a single chain is considered and the lattice spacing is denoted by $a$.

The problem is treated in the half metal limit when the onsite $ |J_0| \gg t$. Since physically $|J_0| < U$ it is also implied that $U \gg t$ and which corresponds to   highly correlated electrons.  This permits{\it (2)\/} a spinless majority electron operator $c^\dagger_{i} \equiv c^\dagger_{i\uparrow}$ to be defined, and in terms of this, the minority $c^\dagger_{i\downarrow} = s^-_i c^\dagger_{i}$ where $s^-_i$ is the conduction electron spin lowering operator.   For large $J_0$ only the largest total angular momentum manifold  is relevant. At a site occupied by a conduction electron, only states with $J=S+1/2$ need to be accounted for and it follows{\it (2)\/} that $\vec s_i = \vec S_i/(2S)$. Physically this implies that the conduction electrons pass adiabatically through the wall, i.e., the system always stays close to the ground state, as discussed in the text. 
This is {\it not\/} an assumption about local equilibrium of the conduction electrons but rather is an immediate and {\it rigorous\/} consequence of the Wigner-Eckart theorem. With this the conduction spin degrees of freedom are sub-summed into those which describe the local moments. In this limit there is a separation of the spin and charge degrees of freedom. The charge sector is spanned by the states created by the spinless fermions $c^\dagger_{i}$, while $\vec S_i$ is the unique spin vector. 

As outlined in the Report, the Holstein-Primakoff transformation $S_{iz}=S-b^\dagger_i b^{}_i$, $S^+_i = (2S- b^\dagger_i b^{}_i)^{1/2}b^{}_i \approx (2S)^{1/2}b^{}_i$ is used to quantize the spins. The local axis of quantization is specified by the Euler angles $\theta_i$ and $\phi_i$. In order that the resulting magnon energies are positive definite, it is necessary that these angles correspond to the  {\it classical\/} equilibrium direction. When substituted into $\cal H$ the coefficient of higher order interactions decreases by a factor of $(2S)^{1/2}$ for each additional $b^\dagger_i$ or  $b^{}_i$. It follows that this approach generates an expansion in $1/S$ about the classical solution.

It is important to understand the role of the contacts in relation to conservation of the angular momentum $S_z$. In the absence of contacts, $[S_z, {\cal H}]=0$, so that $\dot S_z = 0$ and the wall cannot move. It is {\it wrong\/} to conclude that this fact reflects an infinite mass. Observe, when the surface integral $\int  \vec j_s \cdot d \vec A \ne 0$, where $\vec j_s $ is the {\it spin\/} current density, it {\it must\/} be the case that in the local  equation for $d \vec S_i/dt$ there is a $\nabla\cdot  \vec s \propto j_s$,  which couples this current to the spin dynamics. This is the ${\cal H}_{ts}$ discussed in the Report and below. In the half metal limit assumed in this microscopic theory $j_s = j_e$ the charge current density. 

In the absence of a wall, for large $U$, the kinetic energy term is $\hat T = -t\sum_{\langle ij\rangle} (c^\dagger_i c^{}_j+ s^-_i c^\dagger_i c^{}_j s^+_j)+H.c.$ Following the text, the  Bloch wall is generated via a (SU(2)) gauge transformation of the conduction electrons. Since there is a single conduction electron fermion this simplifies to the top entry in the conduction electron spinor:
$$
c^{\dagger}_j \to e^{i\theta_i s_{iy}} c^{\dagger}_j 
\eqno(S2)
$$
Then $e^{i\theta_i s_{iy}} = e^{i(\theta_i/2) \sigma_{iy}} = \cos (\theta_i/2) + i\sin (\theta_i/2)  \sigma_{iy}$ since ${\sigma_{iy}}^2 =1$. Finally $\sigma_{iy} = - i (s^+_i - s^-_i)$ and since $s^+_i c^{\dagger}_j  \equiv 0$ it follows that 
$$
c^\dagger_i \to (\cos [\theta_i/2] + \sin [\theta_i /2]\,s^-_i)  c^\dagger_i.
\eqno(S3)
$$ 
This and its Hermitean conjugate are substituted into the kinetic energy to give:
$$
\hat T \approx -t\sum_{\langle ij\rangle}\cos [(\theta_i- \theta_j)/2] (c^\dagger_i c^{}_j+ s^-_i c^\dagger_i c^{}_j s^+_j)  - i t\sum_{\langle ij\rangle}\sin [(\theta_i- \theta_j)/2]( s^-_i c^\dagger_i c^{}_j - c^\dagger_i c^{}_j s^+_j)+H.c.
\eqno(S4)
$$ 
Charge motion arises principally via ${\cal H}_t  = -t\sum_{\langle ij\rangle}t_{ij} c^\dagger_i c^{}_j +H.c.$ where $t_{ij}=t\cos [(\theta_i- \theta_j)/2]$. The reduction of $t_{ij}$ in the Bloch wall represents a barrier, however this has a height $\sim SA$ which in reality is quite negligible compared to the Fermi energy $E_F$. The solutions of ${\cal H}_t$ are, to a good approximation, plain $\vec k$-states independent of the wall position or its motion. There can be no ``spin-accumulation" of conduction electrons around the wall.

The current-spin interaction arises from the part  of  $\hat T$ which is linear in $s^\pm_i$. Using the  adiabatic condition $\vec s = \vec S/2S$, the relevant interaction is:
$$
{\cal H}_{ts} = 
-{t\over 2(2S)^{1/2}}{\partial \theta_i \over \partial z}a 
\sum_i c^\dagger_i c^{}_{i+1} (b^{}_i - b^{\dagger}_{i+1}) + H.c.,
\eqno(S5)
$$
where $(\theta_i- \theta_j) \approx (\partial \theta_i / \partial z)a$ has been used and which assumes that the wall width $w\gg a$.
Current carrying eigenstates are also eigenstates of $\hat n_{\vec k}= c^\dagger_{\vec k}c^{}_{\vec k}$ the number operator. When acting on such a state the spin-charge interaction reduces to 
$$
{\cal H}_{ts} = 
 -i { \hbar j a^2 \over 2 e S}
  \sum_i { \partial \theta_i \over \partial z}  a  (2S)^{1/2}  (b^{}_i - b^{\dagger}_{i}).
\eqno(S6)
$$
Here $(2S)^{1/2}(1/2i)  (b^{}_i - b^{\dagger}_{i}) \approx  S^\ell_{iy}$ defined to be strictly perpendicular to the {\it instantaneous\/} axis of quantization. This constitutes an explicit derivation of the key Eqn.~3 of the Report albeit for the limit $p=1$.

The, bi-linear in $s^\pm_i$, part of $\hat T$ is also of importance. When acting on eigenstates of ${\cal H}_t$ these lead to a renormalization, $J_{ij} = J^0_{ij} + (x^\prime t/2S^2)$, of the total inter-site exchange coupling $J_{ij}$. The effective concentration $x^\prime = \langle c^\dagger_i c^{}_j \rangle$. One of the interests in the spintronics field is the possibility of exerting forces on, and pinning of, a Bloch wall. The presence of $x^\prime$ in the expression for $J_{ij}$ indicates that this potentially can be modified and modulated dynamically in a field effect device.


\begin{thebibliography}{8}
\bibitem{9} Using that $\langle \vec S_i \rangle \to \vec S_i$ upon dropping the spin wave operators $b^\dagger_{i}$ and $b^{}_{i}$.
\bibitem{10} If $z_0$ is a coordinate, then it is a general result for the classical ground state energy ${\cal V} \{\theta_i,\phi_i\}$, defined earlier in the text, that,  ${\partial {\cal V} \over \partial z_0} = {i \over \hbar} [{\cal H}, \hat p_z ]$  where $\hat p_z$  is the generator of translations and only the non-operator, i.e., $c$-number, part of commutator is retained. This might be applied directly to the wall coordinate $z_0$ where  $\hat p_z$ is given in the text, or for a rotation by $\theta_i$, if $z_0 \to \theta_i$ and $\hat p_z \to S_{iy}/\hbar \approx (\hbar/2i)(2S)^{1/2}(b^\dagger_i - b^{}_i)$.  The proof: the energy change  $\Delta {\cal V} = R^{-1} {\cal H} R - {\cal H}$ where $R = 1 + i(\hat p_{z}/\hbar) \Delta z$, where by definition ${\cal V}$ is the $c$-number part of the right hand side.  This becomes $\Delta {\cal V} = R^{-1}[ {\cal H}, R] = [ {\cal H}, R] $ since $R^{-1} \to 1$ as $\Delta z (\Delta \theta_i)\to 0$. Then $\Delta {\cal V} = (i/\hbar)[ {\cal H}, \hat p_{z}]\Delta z $ to yield the sought after result.
\end{thebibliography}
\end{document}